\def\Tr{{\rm Tr}\,}

\def\be{\begin{equation}}
\def\ee{\end{equation}}
\def\bea{\begin{eqnarray}}
\def\eea{\end{eqnarray}}
\def\bml{\begin{mathletters}}
\def\eml{\end{mathletters}}
\def\b{\bibitem}

\documentclass{jltp}
\usepackage{graphicx} 

\title{Why Quantum Phase Transitions Are Interesting}

\author{D. Belitz$^*$ and T.R. Kirkpatrick$^+$}

\address{$^*$Department of Physics and Materials Science Institute,\\
             University of Oregon, Eugene, OR 97403, USA\\
$^+$Institute for Physical Science and Technology, and Department of Physics,\\
     University of Maryland, College Park, MD 20742}

\runninghead{D. Belitz and T.R. Kirkpatrick}{Are Quantum Phase Transitions 
             Interesting?}

\begin{document}

\maketitle

\begin{abstract}
This paper discusses why the usual notion that quantum phase transitions
can be mapped onto classical phase transitions in a higher dimension,
and that this makes the former uninteresting from a fundamental theoretical
point of view, is in general misleading. It is shown that quantum phase
transitions are often qualitatively different from their classical
counterparts due to (1) long-ranged effective interactions that are
induced by soft modes, and (2) in the presence of quenched disorder,
an extreme anisotropy of space-time. These points are illustrated using
various magnetic phase transitions as examples.

PACS numbers:
\end{abstract}

\section{INTRODUCTION}

Phase transitions that occur in a quantum mechanical system at zero
temperature ($T=0$) as a function of some non-thermal control parameter,
like pressure, or composition, are
called quantum phase transitions. In contrast to their finite-temperature
counterparts, which are often referred to as thermal or classical phase
transitions, the critical fluctuations one has to deal with at zero
temperature are quantum fluctuations rather than thermal ones, and the need
for a quantum mechanical treatment of the relevant statistical mechanics
makes the theoretical description of quantum phase transitions somewhat
different from that of classical ones. Nevertheless, Hertz, in an important 
paper,\cite{Hertz} showed that the basic theoretical concepts that have been 
used to describe and understand thermal transitions, work in the quantum 
case as well. In particular, he demonstrated in detail how to adapt the
Wilsonian renormalization group (RG)\cite{WilsonKogut} 
to quantum phase transitions.
That such an adaptation should be possible had been observed earlier by
Beal-Monod.\cite{BealMonod}

Hertz also was the first to derive a Landau-Ginzburg-Wilson 
(LGW) order parameter functional for a quantum phase transition from a 
microscopic action. The example he used was itinerant electron magnetism.
Starting from the general expression for the partition function,
\be
Z = \Tr e^{-H/k_{B}T}\quad,  
\label{eq:1}
\ee
he obtained a functional integral representation of $Z$. The integration is
with respect to an order parameter field that we generically denote by the
vector field ${\bf M}(x)$, where $x = ({\bf x},\tau)$ comprises the real
space position ${\bf x}$ and the imaginary time $\tau$.
In the case of a ferromagnet, $\bf{M}$ represents 
the fluctuating magnetization. The partition function becomes,
\be
Z = \int D[{\bf M}]\ e^{S[{\bf M}]} = \int D[{\bf M}]\ e^{\int dx\, 
                                   s({\bf M}(x))}\quad.
\label{eq:2}
\end{equation}
Here $S$ is the action, $s$ is the corresponding ``action density'', 
and $\int dx = \int d{\bf x}\int_0^{\beta}d\tau$, with $\beta = 1/k_B T$.
Equation\ (\ref{eq:2}) is formally equivalent to the expression 
for a classical partition function in terms of functional integrals, except 
that instead of an integration over $d$ space dimensions, it involves an 
integration over $d$ space dimensions and one (imaginary) time dimension. 
This coupling between space and time is a fundamental aspect of quantum 
statistical mechanics. Physically, it represents the effects of quantum 
fluctuations. For finite $\beta$, or nonzero temperatures, the time dimension 
is of finite extent, and consequently it makes no difference to the asymptotic
critical behavior, but for $\beta =\infty$ ($T=0$), it seems to suggest
that the quantum critical behavior can be related to the critical behavior
of a classical system in a higher dimension. Indeed, this is just what Hertz
concluded.\cite{Hertz} To state this mapping precisely, we let $r$ be some
appropriate dimensionless distance from the critical point,
and $\xi$ the correlation length which 
diverges at the transition. The critical exponent $\nu$ characterizes the
divergence of $\xi$,
\be
\xi \sim 1/\vert r\vert^{\nu}\quad.  
\label{eq:3}
\ee
As a critical point (quantum or classical) is approached, a relaxation time
diverges as well. We denote this time scale by
$\xi_{\tau}$, and characterize its divergence in terms of that of $\xi$.
This defines the dynamical scaling exponent $z$, 
\be
\xi_{\tau }\sim \xi^z\quad.  
\label{eq:4}
\ee
Hertz used these ideas to suggest that the critical behavior of a quantum
phase transition in $d$-dimensions is identical to the critical behavior of
a classical system in $d_{\rm eff}=d+z$ dimensions. This was a
generalization of an earlier idea by Suzuki\cite{Suzuki} 
who studied specific spin models where $z=1$.

Naively, this seems to imply that quantum phase transitions are not
very interesting from a fundamental theoretical point of view, since 
classical transitions have been long studied, and are quite well understood. 
Further, in bulk systems, $d=3$, the
effective dimension satisfies (for $z\geq 1$) $d_{\rm eff}\geq 4$. Since the
upper critical dimension of many classical phase transitions is $d_c^+ = 4$,
this is consistent with the notion that most quantum phase transitions are
above the upper critical dimension where mean-field theory gives the exact
critical behavior, making the quantum phase transition doubly uninteresting.

In this paper, we point out ways in which these conclusions can break down.
In particular, we first discuss how the coupling between statics and
dynamics that is inherent in quantum statistical mechanics leads to novel
effects not reflected in the simple mapping $d\rightarrow d+z$. 
(Parenthetically, we note that one can {\it always} map a quantum system on
some contrived classical model with one or more long-ranged interactions.
This is not what we mean by `simple' mapping.) Secondly, we
point out that in the presence of quenched disorder there is an extreme
anisotropy between space and time that typically makes the quantum phase
transition qualitatively different than the corresponding classical one.
The choice to focus on these two aspects is somewhat subjective. There are
other ways in which quantum phase transitions become of fundamental interest,
some of which we mention in the Discussion, but we cannot do all of 
these topics justice in the present format.

The plan of this paper is as follows. In the next section, we discuss power-law
correlations that generally exist in both classical and quantum systems,
even away from any critical point. As we will see, the main distinction
between the classical and quantum cases is that in the classical case these
correlations only occur in the time domain, while in the quantum case they
also exist in space. In Section \ref{sec:III} we show how the 
spatial power-law correlations discussed in Section \ref{sec:II} couple 
to, and ultimately determine, the critical behavior at many quantum phase
transitions. In Section \ref{sec:IV} we discuss the case of quenched disorder, 
and in particular explain why quantum transitions in systems with quenched 
disorder are distinct from {\em any} classical transition. We conclude with 
some remarks in Section \ref{sec:V} Among
other things, we point out experimental connections, and additional reasons
why quantum phase transitions are fundamentally interesting.

\section{GENERIC SCALE INVARIANCE IN CLASSICAL AND QUANTUM PHYSICS}
\label{sec:II}

Homogeneous functions, or power laws, of space or time are
characterized by the absence of any intrinsic length or time scales, 
in contrast to, e.g., exponentials. This absence of scales is referred
to as scale invariance. It is well known to occur at
critical points, where the critical modes become soft, which leads to 
power-law correlation functions.\cite{WilsonKogut} 
Critical points are exceptional points in the
phase diagrams of materials, and reaching them requires the fine tuning of
parameter values. What has become clear only more recently is the fact that 
many systems display what is now known as generic scale invariance (GSI),
namely, power law correlation functions in entire regions of parameter space, 
with no fine tuning at all needed to observe them. GSI is caused by soft 
modes that are
not related to critical phenomena, but rather are due to conservation laws,
or possibly Goldstone modes, that are inherent to the system. In recent
years there has been considerable attention to GSI in systems ranging from
classical fluids, to liquid crystal systems, to disordered electron systems,
and to granular, or sandpile, systems.\cite{GSI}

To understand the origin of GSI in classical systems we consider the
case of a simple classical equilibrium fluid. In this system the best
known example of power-law correlations is the temporal decay of the velocity
autocorrelation, $C_{D}(t)= \langle{\bf v}(t)\cdot {\bf v}(0)\rangle_{eq}$, 
where $\langle\ldots\rangle_{eq}$ denotes an equilibrium thermal average.
Its time integral determines the self-diffusion coefficient, $D$, in
$d$-dimensions via
\be
D = \frac{1}{d}\int_{0}^{\infty}dt\ C_{D}(t)\quad.  
\label{eq:5}
\ee
$C_D(t)$ decays only algebraically for long times, 
\be
C_{D}(t>>t_0)\approx c\,(t_{0}/t)^{d/2}\quad,  
\label{eq:6}
\ee
where $t_{0}$ is the mean-free time between collisions, and $c$ is a
positive constant. This slow decay of correlations was first observed in
computer simulations of hard sphere fluids,\cite{AlderWainwright}
and then understood 
theoretically.\cite{ErnstHaugevanLeeuwenDorfmanCohen} 
It turns out that {\em all} time correlations in classical
fluids that are of physical interest decay as power laws, and collectively 
these effects are known as long-time tails (LTT).

The basic physical idea behind the explanation of the LTT phenomena is that
the hydrodynamic modes, which are the slowest decaying modes in noncritical
systems, determine the long-time behavior of all time correlation 
functions, even those for nonhydrodynamic variables. Of
particular importance among the hydrodynamics processes 
are recollision processes, where after a collision,
the two involved particles diffuse away and then meet again and recollide. We
see this in Eq.\ (\ref{eq:6}), the right-hand side of which is proportional to 
the probability that a diffusing particle returns at time $t$ to the point it
started out from at $t=0$. Note that this argument does not necessarily
imply that any particular time correlation function decays as $t^{-d/2}$, 
since it might not couple strongly to this diffusive process. The argument 
does suggest, though, the possibility of power-law decays for generic
correlation functions in systems
with conserved quantities. Finally, it is important to note that even though
time correlations in equilibrium classical fluids do exhibit GSI, spatial
correlations do not; generically, they decay
exponentially on the length scale of a particle diameter. For very
fundamental reasons, dynamical correlations do not couple to static ones in
classical equilibrium systems, and spatial correlations are of short range.

The situation is fundamentally different in quantum systems. Again, time
correlations generically decay as power laws because of coupling to either
hydrodynamic processes, or to dynamical Goldstone mode fluctuations.
Further, in quantum statistical mechanics, statics and dynamics are
fundamentally coupled together. This suggest that the same soft modes that
cause quantum LTTs may lead to power-law correlation functions in space for
the static, or equal-time, correlation functions. This is indeed what 
generically occurs. To be specific, consider a disordered and interacting 
electron system. In this case, it is well known that so-called 
``weak-localization effects''\cite{LeeRama} 
lead to a low-frequency nonanalyticity in the time correlation
function that determines the electrical conductivity, $C_{\sigma}(\omega)$.
In particular, in the low-frequency, long-time domains, and for $2<d<4$,
one finds
\be
C_{\sigma}(\omega)/C_{\sigma}(\omega=0) = 1 + c\,(i\omega)^{(d-2)/2} + \ldots
    \quad,
\label{eq:7}
\ee
\be
C_{\sigma}(t\gg t_{0})/C_{\sigma }(t=0)\approx -c'\,(t_{0}/t)^{d/2}\quad,
\label{eq:8}
\ee
with $c$ and $c'$ positive constants. 
The coupling between statics and dynamics in
quantum systems suggests that static correlation functions such as the spin
susceptibility, $\chi_{s}({\bf q})$, with ${\bf q}$ the wavevector,
will exhibit a nonanalyticity at ${\bf q} = 0$. Since for the
diffusive processes that lead to the Eqs. (\ref{eq:7},\ref{eq:8}), 
frequency and wavenumber scale as $\omega \sim {\bf q}^2$, one expects
\be
\chi_{s}({\bf q}) = c_{0} - c_{d-2}\,\vert{\bf q}\vert^{d-2} 
                          - c_2\,{\bf q}^2 + \ldots,  
\label{eq:9}
\ee
which corresponds to a power-law long-distance spatial decay,
\be
\chi_{s}(\vert{\bf x}\vert\rightarrow\infty)
    \approx c''/\vert{\bf x}\vert^{2(d-1)}\quad.  
\label{eq:10}
\ee
Explicit calculations confirm these results,\cite{LeeRama,us_fm_dirty} 
with positive values for the various constants. All of these
nonanalyticities are properties of the system at $T=0$; at nonzero
temperature $C_{\sigma}$ and $\chi_s$ are analytic functions of
$\omega$ and ${\bf k}$, respectively. This illustrates an important point: 
In addition
to the coupling between statics and dynamics, which induces spatial
long-range correlations in equilibrium quantum systems, there are in
general many more soft modes, which lead to long-range correlations,
at $T=0$ than at $T>0$.

The question then arises: How do these inherent {\em generic} 
power-law correlations
affect, and possibly modify, the critical behavior in the vicinity of
a critical point?
One anticipates dramatic effects from these correlations, if they
couple to the order parameter of the phase transition. This point is
addressed in the next section.

\section{GENERIC SCALE INVARIANCE AND QUANTUM PHASE TRANSITIONS}
\label{sec:III}

We now discuss an example that shows how GSI can dramatically affect
the critical behavior at a quantum phase transition. The example we
choose is that of the quantum ferromagnetic transition in both clean
and disordered itinerant electron
systems. We start with the partition
function, $Z$, for a general disordered electron gas. In a field
theoretic formalism,
\be
Z=\int D[\bar{\psi},\psi]\ e^{S[{\bar\psi},\psi]}\quad.  
\label{eq:11}
\ee
Here the functional integration is with respect to Grassmann valued fields, 
$\bar{\psi}$ and $\psi$, and $S$ is the action. For simplicity we
consider a $d$-dimensional continuum model of interacting electrons.
Of the various electron-electron interaction channels, we pay
particular attention to the particle-hole spin-triplet contribution,
$S_{int}^t$, since this
is what is responsible for ferromagnetism. We denote the coupling constant
in this channel by $\Gamma_t$. Writing only this
term explicitly, and denoting the spin density by ${\bf n}_s$, the
action reads,
\be
S = S_0 + S_{int}^t = S_0 + \frac{\Gamma_t}{2}\int dx\ {\bf n}_s(x)
     \cdot{\bf n}_s(x)\quad.  
\label{eq:12}
\ee
Here $S_{0}$ contains all contributions to the action other than $S_{int}^t$. 
In particular, it contains the effects of disorder, as well as the
particle-hole spin-singlet and the particle-particle interactions. 
Following standard procedure,\cite{Hertz} 
we perform a Hubbard-Stratonovich decoupling of $S_{int}^t$ by introducing 
a classical vector field ${\bf M}(x)$ with components $M^i$ that couples 
linearly to ${\bf n}_s(x)$ and whose average is proportional to the 
magnetization, and we integrate out all fermion degrees of freedom. 
The partition function then reads,
\begin{equation}
Z = e^{-F_0/T}\int D[{\bf M}]\ e^{-\Phi [{\bf M}]}\quad,  
\label{eq:13}
\end{equation}
where $F_{0}$ is the non-critical part of the free energy. Expanded in a power
series, the LGW functional $\Phi$ reads,
\begin{eqnarray}
\Phi[{\bf M}]&=&\frac{1}{2\Gamma_t}\int dxdy\,\delta(x-y)\,
               {\bf M}(x)\cdot{\bf M}(y) + \sum_{n=2}^{\infty}a_n\int
                   dx_1\ldots dx_n
\nonumber\\
&&\times\chi_{i_1\ldots i_n}^{(n)}(x_1,\ldots,x_n)\,
       M^{i_1}(x_{1})\ldots M^{i_n}(x_n)\quad,
\label{eq:14}
\end{eqnarray}
where $a_{n}=(-1)^{n+1}/n!$ . The coefficients $\chi^{(n)}$ in 
Eq. (\ref{eq:14})
are connected $n$-point spin density correlation functions of a reference
ensemble whose action is given by $S_0$. Notice that this reference ensemble
contains the effects of all interaction amplitudes other than $\Gamma_t$,
as well as the effects of disorder.

\subsection{Disordered Ferromagnets}

To be specific, we first consider the disordered case, i.e. itinerant
ferromagnets with quenched, nonmagnetic impurities. To
make the points relevant for this paper we focus on the quadratic or
Gaussian part of the LGW functional, $\Phi_2$, and carry out the 
disorder average by means of the replica trick.\cite{Grinstein}
With $\alpha$ denoting the replica label, and $q_n = ({\bf q},\Omega_n)$ 
a 4-momentum, where $\Omega_n = 2\pi Tn$ is a bosonic Matsubara frequency, 
we have,
\begin{equation}
\Phi_2 = \frac{1}{2}\sum_{q_n}\sum_{\alpha}\left[\frac{1}{\Gamma_t}
         -\chi^{(2)}(q_n)\right]\,\vert {\bf M}^{\alpha}(q_n)\vert^2\quad.
\label{eq:15}
\end{equation}
Here $\chi^{(2)}$ is the Fourier transform of the dynamic spin
susceptibility in the reference ensemble. Spin density conservation implies
that at small frequency and wavenumber, $\chi^{(2)}$ has a diffusive
structure,
\be
\chi^{(2)}(q_n) = \chi_s({\bf q})\,\frac{D{\bf q}^2}{\vert \Omega_n
                   \vert + D{\bf q}^2}\quad,  
\label{eq:16}
\ee
where $D$ and $\chi_{s}({\bf q})$ are the spin diffusion coefficient and
the static spin susceptibility, respectively, in the reference ensemble. 
In the critical limit, where the frequency must be taken to zero before the
wavenumber, we have,
\be
\chi^{(2)}(q_n) = \chi_s({\bf q})\,\left[1 - \vert\Omega_n\vert 
                  /D{\bf q}^2 + \ldots \right]\quad.  
\label{eq:17}
\ee

Now comes the crucial step of calculating $\chi_{s}({\bf q})$ in the
reference ensemble. Until recently it was universally, if tacitly,
assumed that $\chi_s$ was an
analytic function of ${\bf q}^2$. The physical argument behind this
assumption was that the reference ensemble corresponds to 
a physical system that is far from
any critical point. Indeed, the whole philosophy of separating out the
spin-triplet interaction and introducing an order parameter field was
based on the desire to separate the slowly decaying order parameter modes
from all other modes, which were assumed to decay exponentially. 
However, as discussed in Section \ref{sec:II} above, it is now known
that this reasoning is not correct, since any interacting 
itinerant electron system, especially at $T=0$,
has long-range correlations {\em everywhere} in
the phase diagram, even far from any critical point. For small
values of ${\bf q}^2$, for $d>2$, and at $T=0$, the leading
behavior of $\chi_s$ is given by Eq.\ (\ref{eq:9})
where $c_0$, $c_{d-2}$, and $c_2$ are positive constants. 
Notice that the susceptibility
decreases with increasing wavenumber, which is due to the following
effect. The diffusive dynamics of the electrons effectively increases
the strength of the electron-electron interaction compared to a clean
system, which means that disorder increases the homogeneous spin
susceptibility. At nonzero wavenumbers, this effects gets smaller,
which accounts for the negative sign of the nonanalytic term in 
Eq.\ (\ref{eq:9}).
With decreasing spatial dimension, the nonanalyticity becomes stronger 
for phase space reasons, and for $d\leq 2$ the electrons 
get localized\cite{LeeRama} and a different theory is needed.

Using Eqs.\ (\ref{eq:16},\ref{eq:17}) and (\ref{eq:9}) 
in Eq.\ (\ref{eq:15}) yields
\be
\Phi_2 = \frac{1}{2}\sum_{q_n}\sum_{\alpha}\left[r_0 + \vert{\bf q}\vert^{d-2}
         + {\bf q}^2 + \vert\Omega_n\vert/{\bf q}^2\right]\,
         \vert{\bf M}^{\alpha}(q_n)\vert^2\quad.  
\label{eq:19}
\ee
Here
\be
r_0 = 1/\Gamma_t - \chi_{s}({\bf q}=0)\quad,
\label{eq:20}
\end{equation}
is the bare distance from the critical point, and we have omitted various
prefactors in Eq.\ (\ref{eq:19}) which are not essential for our discussion. 
As noted in Section\ \ref{sec:II}, the term
$\vert{\bf q}\vert^{d-2}$ in Eq.\ (\ref{eq:19}) 
implies long-range interactions in real space
between spin density fluctuations. It is well known from the theory of
classical phase transitions that long-range interactions suppress
fluctuations effects, and that the 
critical behavior in systems with such
interactions can usually be determined exactly.\cite{FisherMaNickel} 
For example, simple RG
arguments suggest that all terms higher than $n = 2$ in the Landau
expansion, Eq.\ (\ref{eq:14}), are irrelevant (in the RG
sense) for $d>2$, so that the upper critical dimension
for our phase transition is $d_c^+ = 2$. 
From this it follows that many of the critical exponents
can be determined exactly by considering $\Phi_2$ only.

Although extensive work is necessary to ascertain that the above arguments
are indeed valid, it turns out that several exponents are indeed determined
exactly by $\Phi_2$. Among these are the order parameter susceptibility
exponent $\gamma$, the correlation length exponent $\nu$, the dynamical
scaling exponent $z$, and the exponent $\eta$, which determines the
critical wavenumber dependence of the order parameter susceptibility.
The values of these exponents are, for $2<d<4$,\cite{us_fm_dirty}
\begin{equation}
\gamma = 1\quad,\quad\nu = 1/(d-2)\quad,\quad\eta = 4-d\quad,\quad z = d\quad.
\label{eq:21}
\end{equation}
For $d\geq d_c^{++}= 4$, all of these exponents `lock into' their mean-field
values,\cite{Hertz} $\gamma = 1$, $\nu = 1/2$, $\eta = 0$, and $z = 4$.

To determine the critical exponents $\beta $ and $\delta $ we need the
equation of state in the ordered phase. It has been shown in 
Ref.\ \onlinecite{us_fm_dirty} that the same singularities that lead to the 
term $\vert{\bf q}\vert^{d-2}$ in Eq.\ (\ref{eq:19}) lead to
nonanalyticities in the equation of state. For small values of the
magnetization $m$ one obtains
\be
rm + m^{d/2} + m^{3} = h\quad,  
\label{eq:25}
\ee
with $h$ the magnetic field, and $r$ the physical distance from the critical
point, i.e. the renormalized counterpart of the bare distance $r_0$, 
Eq.\ (\ref{eq:20}). Once again, prefactors have been omitted. 
Notice the $m^{d/2}$
term, which occurs in addition to what is otherwise an ordinary mean-field
equation of state. Its origin are again the effective long-range interactions
between the spin fluctuations. For $d<6$, the $m^{d/2}$ term dominates 
the usual $m^{3}$, and hence determines
the critical exponents $\beta$ and $\delta$. Accordingly, for $2<d<6$,
we have
\begin{equation}
\beta = 2/(d-2)\quad,\quad \delta = d/2\quad.  
\label{eq:26}
\end{equation}
Note that these relations imply yet another upper critical dimension, namely, 
$d_{c}^{+++}=6$, defined as the dimension above which $\beta$ and $\delta$ 
``lock into'' their mean-field values of $1/2$ and $3$, respectively.

We conclude that GSI effects largely determine the quantum critical behavior 
at the ferromagnetic transition in disordered
itinerant electron systems. We also note that the above discussion has been
simplified for pedagogical purposes, and to underscore the relevant physics.
A different approach, that keeps all of the soft modes explicitly
and on equal footing, rather than integrating out all of the fermionic degrees
of freedom, is technically more satisfactory. 
Such a more detailed analysis shows
that complicated logarithmic corrections to scaling occur in all dimensions
$d<d_c^{+++}$, which lead to log-log-normal terms multiplying the power-law
critical behavior characterized by the above exponents.\cite{us_fm_dirty}
However, the basic physical arguments given above are unaffected by this
complication.

\subsection{Clean Ferromagnets}

In the previous subsection, we considered the problem of disordered 
quantum ferromagnets. However, the only point in that discussion where 
the disorder was
important was the diffusive dispersion relation of the ``extra'' (in
addition to the
soft critical mode) soft modes. This raises the possibility that
similar effects might exist in clean itinerant ferromagnets. The first
question that arises in this context is what, if anything, will replace the
$\vert{\bf q}\vert^{d-2}$ term in the static spin susceptibility, 
Eq.\ (\ref{eq:9}), in clean
systems. To answer this, let us consider the perturbation theory for 
$\chi_s({\bf q})$. What leads to the nonanalyticity in Eq.\ (\ref{eq:9}) is the
coupling of two diffusive modes, which mathematically takes the form of a
mode-coupling integral of the type,
\be
\int d{\bf k}\int d\omega\ \frac{1}{\omega + {\bf k}^2}\,
   \frac{1}{\omega + ({\bf k}+{\bf q})^2}\quad,  
\label{eq:28}
\ee
with ${\bf q}$ the external wavenumber. For simplicity, we have set the 
diffusion coefficient equal to unity. RG techniques 
have shown that this indeed gives the leading small wavenumber behavior of 
$\chi_{s}({\bf q})$.\cite{us_fermions}

What changes in a clean system? The soft modes are still the density and
spin density fluctuations, as well as more general particle-hole
excitations. All of these have a linear dispersion relation, i.e., 
$\omega\sim\vert{\bf q}\vert$. One might thus expect $\chi_s({\bf q})$
in a clean system to have a mode-mode coupling contribution analogous to
Eq.\ (\ref{eq:28}), but with ballistic modes instead of diffusive ones,
\be
\int d{\bf k}\int d\omega\ \frac{1}{\omega + \vert{\bf k}\vert}\,
     \frac{1}{\omega + \vert{\bf k} + {\bf q}\vert}\quad.  
\label{eq:29}
\ee
In generic dimensions, expanding in $\vert{\bf q}\vert$ leads to
\be
\chi_s({\bf q})\sim {\rm const} + d_{d-1}\vert{\bf q}\vert^{d-1}
    - d_2\vert{\bf q}\vert^2\quad,  
\label{eq:30}
\ee
at $T=0$. Here $d_{d-1}$ and $d_{2}$ are constant prefactors. For $d<3$,
the nonanalytic term in Eq.\ (\ref{eq:30}) represents the leading small 
wavenumber dependence of $\chi_s$. In $d=3$, one finds a 
${\bf q}^2\,\ln (1/{\bf q}^2)$ term, and in $d>3$ the analytic 
${\bf q}^2$ contribution is the leading one.\cite{us_chi_s} 

The coefficients $d_{d-1}$ and $d_2$ in Eq.\ (\ref{eq:30}) are positive in 
low order perturbation theory with respect to the electron-electron
interaction.\cite{us_chi_s} Physically, the fact that the second term in 
Eq.\ (\ref{eq:30}) is
positive is consistent with the idea that the correlation effects which lead
to this term suppress ferromagnetism. Note that the sign of the nonanalytic
term in Eq.\ (\ref{eq:30}) is opposite that of the one in Eq.\ (\ref{eq:9}).
We further note
that the positive $c_{d-2}$ in Eq.\ (\ref{eq:9}) leads to the 
positive $\vert{\bf q}\vert^{d-2}$ in Eq.\ (\ref{eq:19}) and the positive 
$m^{d/2}$ in Eq.\ (\ref{eq:25}). This implies, as has been confirmed by
explicit calculations,\cite{us_fm_clean}
that the equation of state in a clean itinerant ferromagnet at $T=0$ is,
\be
rm - m^d + m^3 = h\quad.  
\label{eq:31}
\ee
For $d\leq 3$ this equation of state predicts a discontinuous, or first
order, ferromagnetic transition at zero temperature.

\begin{figure}
\centerline{\includegraphics[width=5.0in]{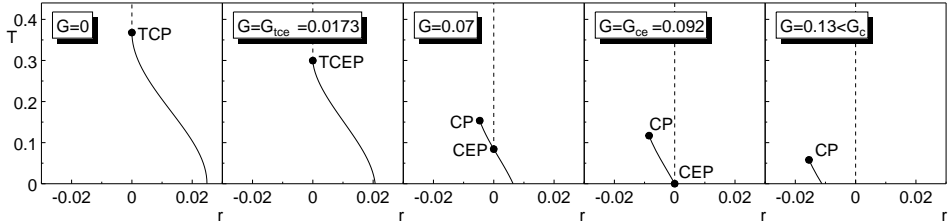}}
\caption{Phase diagrams for various strengths of the disorder $G$ showing
a tricritical point (TCP), critical end points (CE), and critical points (CP).
Solid lines denote first order transitions, and dashed lines second order ones.
From Ref.\ \onlinecite{us_first_order}.}
\label{fig:1}
\end{figure}

More detailed calculations confirm this result.\cite{us_first_order}
Further, they yield a
tricritical point at low temperatures. The net result is that in very clean
systems the magnetic phase transition should be generically of first order.
For sufficiently disordered systems, the transition is continuous, and the
critical behavior is given by the results discussed in the previous
subsection. Some of the typical
phase diagrams in the $T$-$r$ plane, for various disorder strengths, are 
shown in Fig.\ \ref{fig:1}.
As in the disordered case, we again conclude that for clean ferromagnets,
GSI effects largely determine the nature of the quantum phase transition in
systems with physical values of the spatial dimensionality.

\section{ANISOTROPIC QUANTUM PHASE TRANSITIONS: THE CASE OF QUENCHED DISORDER}
\label{sec:IV}

Apart from their tendency to be influenced by GSI phenomena, there
is another aspect of quantum phase transitions that makes them 
fundamentally different from classical ones. To understand this
point, it is useful to consider a classical transition in an anisotropic 
system, i.e. one where all of the $d$ spatial dimensions are not equivalent.
As long as the range of interactions in
all directions is finite, the difference between the
different directions becomes irrelevant, in the RG sense,
as the critical point is approached. This makes sense physically: As one
considers longer and longer length scales, all interactions effective become
short-ranged, unless they decay slower than $1/\vert{\bf x}\vert^{d+2}$
in a particular direction. However, the situation is different if the
interactions in some dimensions are of long, or possibly infinite, range.

Quantum phase transitions with quenched disorder are generic examples where
interactions in one direction (viz., the time direction) are of infinite range
while the interactions in the other directions (viz., the spatial ones) are
of finite range. To see this, consider
a LGW action $S[{\bf N}]$ for, say, the quantum antiferromagnetic phase
transition in an itinerant electron system with nonmagnetic quenched
disorder, with ${\bf N}$ the fluctuating staggered magnetization:
\begin{eqnarray}
S[{\bf N}]&=&\frac{1}{2}\int d{\bf x}\int_{0}^{1/T}d\tau\ 
             {\bf N}({\bf x},\tau)\,
             [r_0 + \delta r({\bf x}) - \nabla^2 + \partial_{\tau}^2]\,
             {\bf N}({\bf x},\tau)
\nonumber\\
&&\hskip 30pt + u\int d{\bf x}\int_{0}^{1/T}d\tau\ 
   \left({\bf N}({\bf x},\tau)\cdot{\bf N}({\bf x},\tau)\right)^2\quad,
\label{eq:32}
\end{eqnarray}
with $u$ a positive constant, and $\delta r({\bf x})$ a random-mass term
that represents the effects of nonmagnetic impurities. For simplicity,
we take it to be delta-correlated in space,
\be
\{\delta r({\bf x})\delta r({\bf y})\}_{\rm dis} = G\,\delta({\bf x} -
    {\bf y})\quad.
\label{eq:33}
\ee
Here $\{\ldots\}_{\rm dis}$ denotes the disorder average, and $G$ is
a measure of the disorder strength.
We can now see the extreme anisotropy inherent in
quantum statistical mechanics: 
The disorder is delta-correlated in space, but,
formally, of infinite range in time. Using replica methods to integrate out
the disorder leads to an additional quartic term with one spatial integral,
but two time integrals, due to this infinite-range interaction in the time
domain.

The sign of this new quartic term is the opposite of that of the $u$-term in
Eq.\ (\ref{eq:32}). Further, the extra imaginary time integral in the new
term, which is due to the long-range
interaction in time, means that compared to the original
quartic term, it is of order $O(G/T)$ and thus diverges as
temperature goes to zero. Together with the negative sign, 
this suggests a tendency for
local ${\bf N}$-ordering even for $r>0$. That is, instanton effects (i.e., the
formation of droplets)
are much more important at quantum phase transitions than they are at the
corresponding classical transition.

Technically, a perturbative RG solution of the field theory,
Eq.\ (\ref{eq:32}), shows that the
disorder scales to infinity, consistent with the suggestion that the
disorder term is very important.\cite{us_afm} A functional RG approach
that treats the disorder nonperturbatively has been developed by Fisher 
for various one-dimensional models.\cite{DSF} 
He, and others, have shown that disorder, at least
in one-dimension, leads to a number of novel features of quantum phases, and
quantum phase transitions. In particular, it was shown that even in the
disordered phase ($r<r_c$) a number of so-called Griffith's phase 
effects\cite{Griffiths} occur. For
example, for $0<r<r_{c1}<r_c$ the magnetic susceptibility becomes singular even
in the absence of long-range order. At other `critical' values
of $r$, which are located in the interval $r_{c1}<r<r_c$,
higher order susceptibilities become singular. Such effects are
not seen in classical systems with interactions of finite range. Further,
for some model systems it
has been established that the quantum phase transition at $r=r_c$,
where true long-range order sets in, is characterized by exponential
singularities rather than by power laws.\cite{Villain,DSF} 
This behavior is unlike that at standard classical phase transitions, and
there is extensive numerical evidence that is consistent with this
theoretical picture.\cite{APY}

Finally we note that the quartic term of long range in time discussed above
also occurs at the
disordered ferromagnetic transition discussed in Section \ref{sec:III}
There, it has no effect on the leading critical behavior because the 
long-range spatial fluctuations due to the extra soft
modes effectively suppress this disorder term. Physically, the interactions
extend over a large enough region to smooth out the remaining disorder
effects.

\section{DISCUSSION}
\label{sec:V}

In this paper we have shown that in contrast to popular lore, in many
respects quantum phase transitions are fundamentally different from standard
classical phase transitions and therefore interesting from a basic
statistical mechanics viewpoint. We have focused on two different aspects.
We first discussed how the coupling between statics and dynamics in quantum
statistical mechanics generically leads to long-range interactions between the
order parameter fluctuations at quantum phase transitions. This
source of long-range interactions is dynamical in nature and is therefore
absent in classical systems, where the statics and dynamics do not couple. 
We then discussed how, in the presence of quenched disorder, any action
describing a quantum phase transition becomes extremely anisotropic. 
This effect, too,
can be viewed as a type of long-range interaction, but in this case the
long range occurs in the time domain.
All of the consequences of this last effect, in particular, are not yet
fully understood.

There is some existing experimental support for these effects.
First, in very clean bulk itinerant electron systems that order
ferromagnetically, the transition does appear to become first order, or
discontinuous, at low temperatures. This has been observed in both 
MnSi\cite{Pfleiderer} and UGe$_{2}$.\cite{Saxena} 
Further, in MnSi the transition becomes continuous if disorder is 
introduced into the systems, in accord with the theoretical expectations. 
ZnZr is another system where the magnetic transition temperature is very low, 
but a continuous transition is observed.\cite{Pfleiderer} 
However, in this case it is not clear
whether the samples studied have been
sufficiently clean for the transition to be discontinuous. The precise
critical behavior at the continuous transitions in the disordered case 
has not been studied so far. Such experiments would be of great interest.

From a theoretical point of view, it is interesting that there are 
classical analogs of the two effects we have discussed.
A simple classical analog of the first one is an equilibrium
system in a phase with Goldstone modes that couple strongly to an order
parameter undergoing a phase transition. An example is the thermal
ferromagnetic transition in compressible magnets.\cite{Aharony}
However, in order to get this effect
more generically one needs a coupling between statics and dynamics.
While this does not
occur in classical equilibrium systems, it does in classical nonequilibrium
systems. It has been known for some time that these systems
in general exhibit generic
scale invariance,\cite{GSI} and that the order parameters
for phase transitions in such systems
couple to these long-range correlations. The first study of a transition of
this type considered phase separation in a binary liquid under
shear.\cite{Kawasaki}

It should also be pointed out that there are other cases where a
quantum phase transition is
either distinct from any classical transition, and therefore of significant
theoretical interest, or where the corresponding classical transition
has not been studied, and hence the universality class is not known. 
The first category is the case of
spin chains, where half-integer spin systems are fundamentally different
than integer spin ones. It is known, for example, that half-integer
antiferromagnetic spin chains have a critical phase while integer spin
chains are always disordered.\cite{Haldane} 
This effect has no simple classical analog. An
example of the case where the corresponding classical transition has not
been studied, are the phase transitions between quantum Hall 
states.\cite{Sondhi_etal} It is also worth mentioning that there are
speculations that an underlying quantum phase transition might be
responsible for the apparent non-Fermi liquid behavior observed in
high-temperature superconductors.\cite{SantaBarbara}

Finally, it should be pointed out that there are indeed some quantum phase
transitions that can be mapped onto standard quantum phase transitions in
$d+z$ dimensions, i.e. where the `simple' mapping, as defined in the
Introduction, works. An example is the antiferromagnetic transition in clean
insulators or itinerant electron systems in $d>1$. In
general it appears that this `simple' mapping is possible if the system is
clean, and any soft modes do not significantly couple to the order parameter
fluctuations.

\section*{ACKNOWLEDGMENTS}

It is with great pleasure that we dedicate this paper to Professor Peter
W{\"o}lfle on the occasion of his 60th birthday. This work was supported by the
NSF under grant Nos. DMR-98-70597 and DMR-99-75259.

\end{document}